\documentclass[twocolumn,superscriptaddress,showpacs,pra]{revtex4}
\usepackage{epsfig}
\usepackage[bookmarks=false]{hyperref}
\usepackage{delarray}
\usepackage{amsmath, amssymb}
\usepackage{bm}

\begin{document}
\title{Early Detection of Rogue Waves by the Wavelet Transforms\footnote{This article can be cited as: Bayindir, C., 2016, ``Early detection of rogue waves by the wavelet transforms", Physics Letters A (380), pp. 156-161.}}

\author{Cihan Bayindir}
\email{cihan.bayindir@isikun.edu.tr}
\affiliation{Department of Civil Engineering, Isik University, Istanbul, Turkey}

\begin{abstract}
We discuss the possible advantages of using the wavelet transform over the Fourier transform for the early detection of rogue waves. We show that the triangular wavelet spectra of the rogue waves can be detected at early stages of the development of rogue waves in a chaotic wave field. Compared to the Fourier spectra, the wavelet spectra is capable of detecting not only the emergence of a rogue wave but also its possible spatial (or temporal) location. Due to this fact, wavelet transform is also capable of predicting the characteristic distances between successive rogue waves. Therefore multiple simultaneous breaking of the successive rogue waves on ships or on the offshore structures can be predicted and avoided by smart designs and operations.

\end{abstract}
\maketitle

\section{Introduction}
Rogue waves, waves with a height more than 2-2.2 times the significant wave height in the wave field \cite{Kharif}, present a danger to life and their result can be catastrophic and costly. The early detection of rogue waves in the chaotic ocean is a must for the early-warning systems that ensure the safety of the marine travel and the offshore structures in stormy conditions. However the early detection of rogue waves in an extremely hard problem. First of all rogue waves appear in stormy conditions therefore accurate prediction of weather conditions is a must. Secondly rogue waves appear and disappear in a length (time) scale that are on the order of their width \cite{Akhmediev2011}. Due to the rapidly changing nature of the rogue waves, the reliability of the forecast of the rogue waves at the current stage is not high and it is hard to expect that it will become more reliable in the near future \cite{Akhmediev2011}.

This vital problem has been disregarded for a long time and has only been studied for almost a decade \cite{Akhmediev2011, Islas}. The proposed approach in \cite{Akhmediev2011} is to continuously measure the part of the whole surface spectrum in real time and use the triangular Fourier spectra of the growing rogue waves in early stages of their development in a chaotic wave field before the peak appears \cite{Akhmediev2011}. 

In this paper, we show that using the wavelet transforms for the early detection of rogue waves over or with the Fourier transforms is more advantageous than using the Fourier transforms solely. With this motivation, we use a simple model similar to the one discussed in \cite{Akhmediev2011}. In our model the complications due to the two-dimensional effects, finite water depth, higher-order dispersion and  higher-order nonlinearity and other factors that exist in reality are ignored \cite{Akhmediev2011}. We numerically generate a chaotic wave field and analyze the wavelet spectra before and after the rogue wave formation. We show that the wavelet spectra exhibits V-shaped high energy regions when high waves are formed. In the case of a rogue wave occurrence we see that at very low scales of the wavelet transform there is energy. This property can be used for the early detection of the rogue waves. Additionally using the numerical model we show that wavelet analysis is also capable of finding the possible location of a rogue wave thus characteristic distances between successive rogue waves (such as the famous ``three-sisters'') can be predicted as well. Therefore wavelet transforms are more advantageous over the Fourier transforms for the early detection of rogue waves when they are used in a model or in the measurements. Using this result, multiple simultaneous breaking of the successive rogue waves on ships or on offshore structures can be predicted. So by means of smart designs and operations, the safety of the ocean travel and offshore operations can be enhanced.

\section{Generation of the Rogue Waves in a Chaotic Ocean Wave Field}

\subsection{Review of the Nonlinear Schr\"{o}dinger Equation and Peregrine Soliton}

\noindent Dynamics of weakly nonlinear deep water ocean waves are described by the nonlinear Schr\"{o}dinger equation (NLSE) \cite{Zakharov1968, Zakharov1972}. One of the most common forms of the NLSE is given by
\begin{equation}
i\psi_t + \frac{1}{2} \psi_{xx} +  \left|\psi \right|^2 \psi =0
\label{eq01}
\end{equation}
where $x,t$ is the spatial and temporal variables, $i$ denotes the imaginary number and $\psi$ is complex amplitude. This notation is mainly used in ocean wave theory whereas $t$ and $x$ axes are switched in fiber optic studies.  NLSE is also widely used in other branches of the applied sciences and engineering to describe various phenomena including but not limited to pulse propagation in optical fibers and quantum state of a physical system. Integrability of the NLSE is studied extensively within last forty years and some exact solutions of the NLSE are derived. Some rational soliton solutions of the NLSE are derived as well. One of the most early forms of the rational soliton solution of the NLSE is the Peregrine soliton \cite{Peregrine}. It is given by
\begin{equation}
\psi_1=\left[1-4\frac{1+2it}{1+4x^2+4t^2}  \right] \exp{[it]}
\label{eq02}
\end{equation}
where $t$ is the time and $x$ is the space parameter. It is shown that Peregrine soliton is a first order rational soliton solution of the NLSE and the higher order rational solutions of the NLSE and a hierarchy of obtaining those rational solutions based on Darboux transformations are given in \cite{Akhmediev2009b}. Details of the Darboux transformations can be seen in \cite{Matveev}. It has been confirmed that the hydrodynamic rogue waves are in the Peregrine soliton form \cite{Chabchoub}. Additionally, throughout many simulations it has been confirmed that rogue waves obtained by numerical techniques which solve the NLSE are in the forms of these first (Peregrine) and higher order rational solutions of the NLSE \cite{Akhmediev2011, Akhmediev2009b, Akhmediev2009a}. 

One of the major problems with the early detection of the rogue waves is their rapidly changing nature. In the literature rogue waves are described as the waves that appear from nowhere and disappear without a trace. A recent analysis given in \cite{Birkholz} which employs the Grassberger-Procaccia nonlinear time series algorithm opposes this description and suggests that early warning times for rogue waves can be enhanced. However as explained in \cite{Birkholz}, at best one may expect to predict an ocean rogue wave a few ten seconds before impact with the current understanding of the rogue wave dynamics. In order to illustrate the rapidly changing nature of the rogue waves in the physical space we give the following numerical example. Consider the Peregrine soliton given in (\ref{eq02}) and the scale, phase and Galilean transformations \cite{dubard2013multi}
\begin{equation}
\psi(x,t) \rightarrow B \psi(Bx,B^2t), \ \ \ B \in \Re^+ 
\label{eq03}
\end{equation}

\begin{equation}
\psi(x,t)  \rightarrow  \exp{[ic]} \psi(x,t), \ \ \ c \in \Re 
\label{eq04}
\end{equation}

\begin{equation}
\psi(x,t) \rightarrow  \psi(x-Vt,t) \exp{[iVx-iV^2t/2]}, \ \ \ V \in \Re
\label{eq05}
\end{equation}

\begin{figure}[h]
\begin{center}
   \includegraphics[width=3.4in]{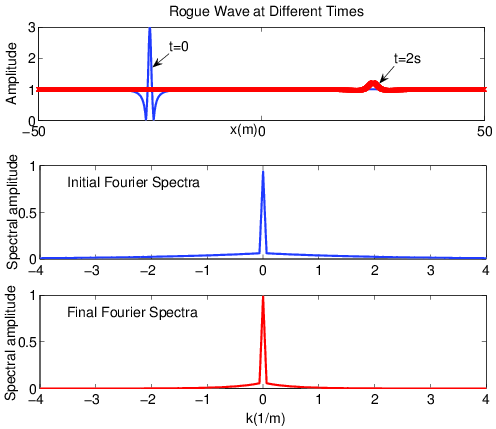}
  \end{center}
\caption{\small Temporal evolution of a rogue wave and its Fourier transform.}
  \label{fig1}
\end{figure}

\begin{figure}[h]
\begin{center}
   \includegraphics[width=3.4in]{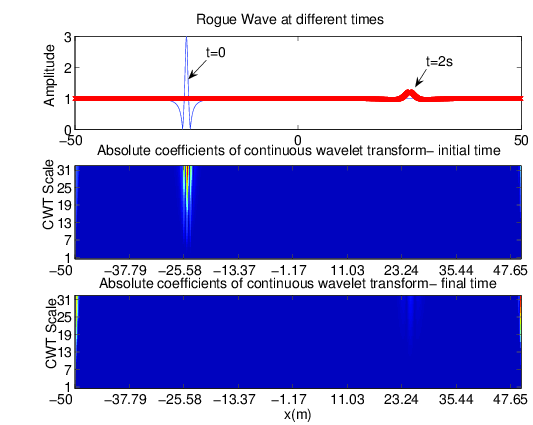}
  \end{center}
\caption{\small Temporal evolution of a rogue wave and its wavelet transform.}
  \label{fig2}
\end{figure}

We obtain a progressive Peregrine soliton by using (\ref{eq02}) and (\ref{eq05}) together and we set $V=25m/s$. Initial profile at $t=0s$ and the final profile obtained at $t=2s$ of time stepping is compared in the Figure~\ref{fig1} below. As depicted in the Figure~\ref{fig1} below, within $2s$ of time the Peregrine soliton almost vanishes.

The Fourier transform of the Peregrine soliton can analytically be calculated as 
\begin{equation}
F(k,t)=\frac{1}{\sqrt{2 \pi}}\int_{-\infty}^{\infty} \psi(t,x)e^ {ikx}dx
\label{eq06}
\end{equation}
which leads to
\begin{equation}
\begin{split}
F(k,t)=& \left[ \frac{1+2it}{\sqrt{1+4t^2}}\exp{\left(-\frac{\left|k\right|}{\sqrt{2}} \sqrt{1+4t^2}  \right)}-\delta(k)  \right]  \\
& .\sqrt{2 \pi} \exp{[it]}
\label{eq07}
\end{split}
\end{equation}
where $k$ is the wavenumber parameter and $\delta$ is the Dirac-delta function \cite{Akhmediev2011}. In the Figure~\ref{fig1} above, we also present the Fourier transforms of the Peregrine solitons in the initial and final times. As discussed in \cite{Akhmediev2011}, the triangular shape of the Fourier spectra can be used for the early detection of the rogue waves.  This is also validated for a chaotic wave field in \cite{Akhmediev2011}.

We propose that using the wavelet transforms in a similar fashion can be more advantageous for the early detection of rogue waves. It is known that wavelet transforms preserve the temporal variations when they are used for calculating the spectra of a time series \cite{Akansu, Chui, Meyer}. Therefore they are capable of finding the spatial (temporal) location of the changes in the wavenumber (frequency) \cite{Akansu, Chui, Meyer}. Therefore wavelet transforms can detect not only if a rogue wave will develop but also where it will develop in the spatial (or temporal) domain. 

 There are many different wavelets such as symlet, Daubechies, coiflet, biorthogonal, Meyer, Haar etc. just to name a few. Depending on the mother wavelet function it may or may not be possible to calculate the wavelet transform of the Peregrine soliton analytically. For illustrative purposes we only present numerical results with symlet wavelet of order 2. We show the wavelet transform of the Peregrine rogue wave propagating on the still water in the Figure~\ref{fig2} below. One can realize the V-shaped high energy region due to the Peregrine soliton. However as Peregrine soliton progresses, its amplitude decays significantly therefore energy in the wavelet spectra reduces as well. It is important to note that smaller the wavelet transform scale, the wavelet is more compressed. Therefore when a dangerous peak due to a rogue wave occurs, wavelet spectra exhibits energy in the lowest scales since a peak is narrow in width. As one can realize from the Figure~\ref{fig1} and Figure~\ref{fig2}, using the wavelet spectra we can identify the occurrence as well as the location of the peak of the rogue well whereas using the Fourier spectra we can not say where
it will appear.

\begin{figure}[htb!]
\begin{center}
   \includegraphics[width=3.4in]{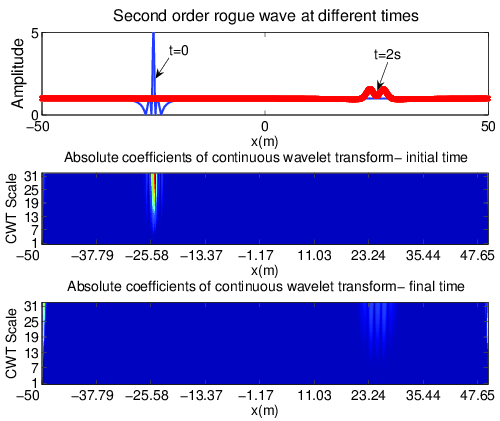}
  \end{center}
\caption{\small Temporal evolution of a second order rogue wave and its wavelet transform.}
  \label{fig3}
\end{figure}

\begin{figure}[htb!]
\begin{center}
   \includegraphics[width=3.4in]{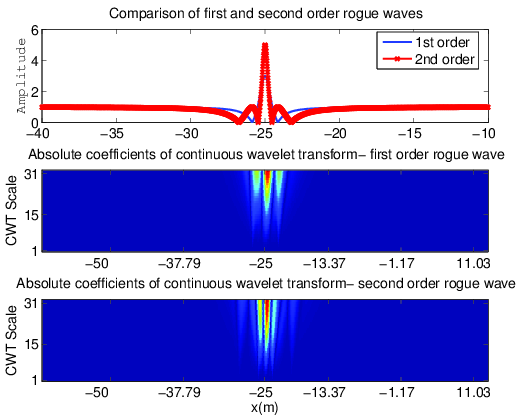}
  \end{center}
\caption{\small Comparisons of wavelet spectra of the first and second order rogue waves.}
  \label{fig4}
\end{figure}

It is known that rogue waves in a wave field may be in the form of first, second or higher order rational solutions of the NLSE \cite{Akhmediev2009b}. A question of practical importance to be discussed is the behavior of the wavelet spectra of second order rogue waves. The form of the second order rogue wave which satisfies the NLSE exactly is given as \cite{Akhmediev2009b}
\begin{equation}
\psi_2=\left[1+\frac{G_2+it H_2}{D_2}  \right] \exp{[it]}
\label{eq071}
\end{equation}
where
\begin{equation}
G_2=\frac{3}{8}-3x^2-2x^4-9t^2-10t^4-12x^2t^2
\label{eq072}
\end{equation}
\begin{equation}
H_2=\frac{15}{4}+6x^2-4x^4-2t^2-4t^4-8x^2t^2
\label{eq073}
\end{equation}
and
\begin{equation}
\begin{split}
D_2=\frac{1}{8} [ \frac{3}{4} & + 9x^2+4x^4+\frac{16}{3}x^6+33t^2 \\
&  +36t^4+\frac{16}{3}t^6-24x^2t^2+16x^4t^2+16x^2t^4 ]
\label{eq074}
\end{split}
\end{equation}

The Fourier spectra of the first, second and higher order rogue waves are compared and discussed in \cite{Akhmediev2011b} in detail where some analytical expressions and mainly numerical and illustrative results are presented. Similar to the first order rogue wave, the second order rogue wave has roughly a triangular Fourier spectra \cite{Akhmediev2011b} when the dirac delta peak due to constant term is ignored. However compared to the Fourier spectra of the first order rogue wave, the Fourier spectra of the second order rogue wave exhibits two dips due to increased number of sidebands in the wave profile.

A similar difference can be observed between the wavelet spectra of the first and second order rogue waves on the Figures~\ref{fig3}-\ref{fig4} below where we have calculated the wavelet spectra numerically. It can be easily realized that both the first and the second order rogue waves have a well defined V-shaped high energy region in their wavelet spectra. This is expected since energy is constrained in a limited region around peaks for both of the profiles.  Therefore detection of V-shaped high energy region can also be used for the detection of the emergence of a second order rogue. The width of the V-shaped high energy region is almost identical however since the second order rogue wave is more peaked, there is energy at the lower scales of the wavelet spectra compared to the wavelet spectra of the first order rogue wave thus the tip of the V-shaped high energy region is better developed. Additionally since the second order rogue wave has an increased number of sidebands, the wavelet spectra exhibits dips in the V-shaped region so that this V-shaped high energy region is has sub-regions. For the early warning or post disaster assessment studies this property can be used to detect which type of rogue wave will develop. However in a chaotic wave field the behavior will be quite complicated. One possible way is that once the emergence of a rogue wave is detected, its wavelet spectra can be compared with the wavelet spectra presented in the Figures~\ref{fig3}-\ref{fig4}. This complicated problem needs to be analyzed in future studies. A similar difference can be observed between the wavelet spectrum of the first, second and higher order rogue waves discussed in \cite{Akhmediev2009b}, more sidebands in the wave profile causes more dips and more sub-regions in the V-shaped region of the wavelet spectra.

A chaotic sea state involves a full range of the wave spectrum. Therefore for the practical purposes it is important to see how the wavelet analysis is going to behave in a chaotic sea state. For this reason we generate a chaotic wave field in the next section using a similar technique to the one discussed in \cite{Akhmediev2011, Akhmediev2009b, Akhmediev2009a}.

\subsection{Initialization of the chaotic wave field} 

We start the rogue wave simulations using a constant amplitude wave with an additive small chaotic perturbation. Such a state is unstable and it evolves into a full-scale chaotic wave field as shown in the numerical simulations described in \cite{Akhmediev2009b, Akhmediev2009a}. The chaotic wave field with this starter evolves into a wave field which exhibits many amplitude peaks, with some of them becoming rogue waves \cite{Akhmediev2011}. In order to model such a chaotic wave field we introduce the initial condition
\begin{equation}
\psi (x,t=0)=1+\mu a(x)\exp{[i\theta]}
\label{eq08}
\end{equation}
where $a(x)$ is a normalized normally distributed random real function with values in interval $[-1,1]$, $\theta$ is a normally distributed random real function with values in the interval $[0,2 \pi]$. The actual water surface fluctuation for this initial condition would be given by $ \left|\psi\right| \exp{[i\omega t]}$ where $\omega$ is the carrier wave frequency.  A similar but different initial condition is described in \cite{Akhmediev2011, Akhmediev2009b, Akhmediev2009a}. Following \cite{Akhmediev2011}, a value of $\mu=0.6$ is selected. It is possible to add perturbations with a characteristic wavelength scale $k_{pert}$, by multiplying the second term in the (\ref{eq08}) by a factor of $\exp(ik_{pert}x)$. Or one can add perturbations with different wavelength scales using Fourier analysis. However in the present study for illustrative purposes we do not consider such as scale.

\subsection{Review of the Split-Step Fourier Method for Numerical Solution of the NLSE} 

Spectral methods is one of the very widely used class of numerical solutions in computational mathematics. Some of their applications can be seen in \cite{bay2009, bayindir2015d, Karjadi2010} and more detailed discussions can be seen in \cite{Canuto, trefethen}. In spectral methods the spatial derivatives are calculated by the orthogonal transforms. The most popular choice for the periodic domain orthogonal transform is the Fourier transform. The time integration is performed using schemes such as Adams-Bashforth and Runge-Kutta etc. \cite{Canuto,  Karjadi2012, Demiray2015}. 

One of the very popular Fourier spectral methods with efficient time integration is the split-step Fourier method (SSFM) which was originally proposed in \cite{Hardin73}. Since then, researchers have proposed many different versions of the SSFM \cite{Bogomolov}. In the SSFM, the time integration is performed by time stepping of the exponential function for an equation which includes a first order time derivative \cite{Taha84}. SSFM is based on the idea of splitting the equation into two parts, the nonlinear and the linear part. For the NLSE, the advance in time due to nonlinear part can be written as
\begin{equation}
i\psi_t= -\left| \psi \right|^2\psi
\label{eq09}
\end{equation}
which can be exactly solved as
\begin{equation}
\tilde{\psi}(x,t_0+\Delta t)=e^{i\left| \psi(x,t_0)\right|^2\Delta t}\ \psi(x,t_0)
\label{eq10}
\end{equation}
where $\Delta t$ is the time step. The linear part of the NLSE can be written as
\begin{equation}
i\psi_t=-\frac{1}{2}\psi_{xx}
\label{eq11}
\end{equation}
Using the Fourier series it is possible to write that
 \begin{equation}
\psi(x,t_0+\Delta t)=F^{-1} \left[e^{-ik^2\Delta t/2}F[\tilde{\psi}(x,t_0+\Delta t) ] \right]
\label{eq12}
\end{equation}
where $k$ is the Fourier transform parameter. Therefore combining (\ref{eq10}) and (\ref{eq12}), the complete form of the SSFM can be written as
 \begin{equation}
\psi(x,t_0+\Delta t)=F^{-1} \left[e^{-ik^2\Delta t/2}F[ e^{i\left| \psi(x,t_0) \right|^2\Delta t}\ \psi(x,t_0) ] \right]
\label{eq13}
\end{equation}
Starting from the chaotic initial condition described above by (\ref{eq08}), the numerical solution of the NLSE equation is obtained for later times by the SSFM. This form of the SSFM requires two FFTs per time step. The number of spectral components are selected as $N=4096$ in order to make use of the fast Fourier transforms efficiently. The time step is selected as $dt=0.04$ which does not cause any stability problems.

\section{Results and Discussion}

In the Figures~\ref{fig5}-\ref{fig10} below, the typical results of the numerical simulation of the chaotic wave field are presented. The actual spatial domain of the numerical simulations is much longer than the one presented in these figures and it is selected as $L=[-500,500]$. It has been previously shown for the similar simulations that, the triangular Fourier spectra begins to develop in a length (time) scale on the order of the width of the rogue wave itself \cite{Akhmediev2011}. We have observed the similar behavior with wavelets, V-shaped high energy regions begin to develop in a time scale on the order of the width of the rogue wave. Although this time scale is too short for early detection, it may still be useful to understand the nature and occurrence of rogue waves \cite{Akhmediev2011}.   

In the Figures~\ref{fig5}-\ref{fig10}, we present the chaotic wave field at the times of $t=40.56s,\ 41.56s,\ 42.36s$, respectively. It is important to note that when t gets bigger and becomes $t=42.56$, the rogue wave will appear at $x=9.52m$ as shown in the Figure~\ref{fig8}. In the Figures~\ref{fig5}-\ref{fig7},  it is possible to realize that V-shaped high energy region is developed around $x=9.52m$ which may be used for the prediction of occurrence and location of a rogue wave.


\begin{figure}[htb!]
\begin{center}
   \includegraphics[width=3.4in]{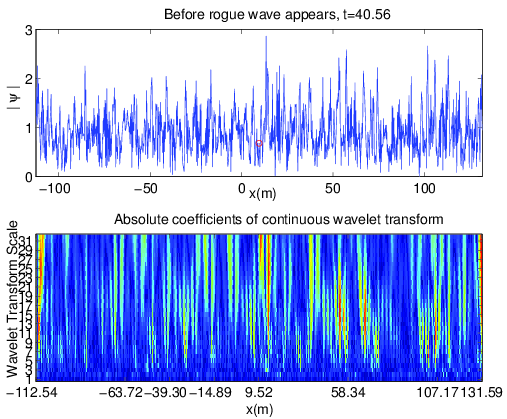}
  \end{center}
\caption{\small  a) Chaotic wave field before rogue wave appears b)  its wavelet transform, t=40.56s.}
  \label{fig5}
\end{figure}

\begin{figure}[htb!]
\begin{center}
   \includegraphics[width=3.4in]{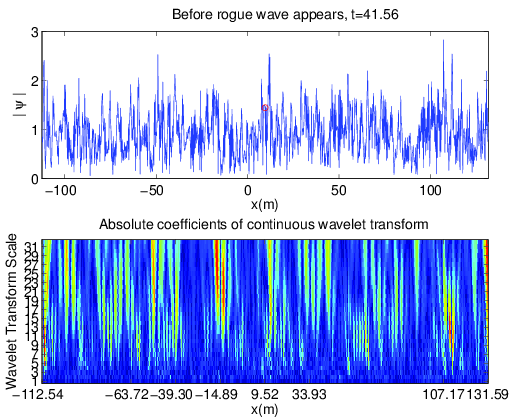}
  \end{center}
\caption{\small a) Chaotic wave field before rogue wave appears b) its wavelet transform, t=41.56s.}
  \label{fig6}
\end{figure}

\begin{figure}[htb!]
\begin{center}
   \includegraphics[width=3.4in]{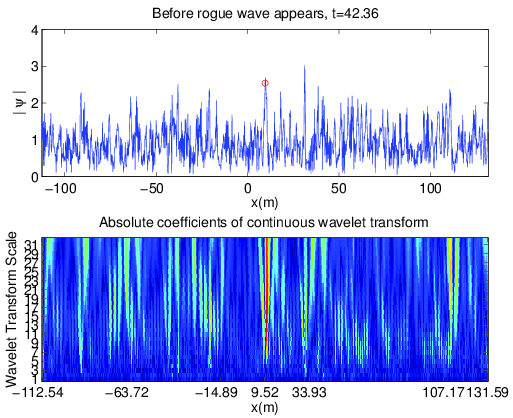}
  \end{center}
\caption{\small a) Chaotic wave field before rogue wave appears b) its wavelet transform , t=42.36s.}
  \label{fig7}
\end{figure}

It is known that as the scale of the wavelet transforms becomes smaller, the wavelet becomes more compressed. Since a rogue wave has a high peak and narrow width, the small wavelet transform scales of the chaotic field should include energy when a rogue wave is about to develop. Therefore we are particularly interested in well-defined V-shaped regions in the wavelet spectra which include some energy in the lowest scales of the wavelet spectra. In the Figure~\ref{fig5} three such regions can be seen around $x\approx 10m$, $x\approx 59m$ and $x\approx 107m$. Therefore in the chaotic field successive high waves may develop  with characteristic distances of  $d \approx 25m$ and exact multiples of this value. Similarly, in the Figure~\ref{fig6}, two such regions can be seen around $x\approx 10m$ and $x\approx 107m$ and in the Figure~\ref{fig7}, two such regions can be seen around $x\approx 9m$ and $x\approx 34m$ which lead to the similar characteristic distance between successive rogue waves for the given chaotic field.

In the Figure~\ref{fig8}, it is observed that the rogue wave with an amplitude close to $5$ has occurred. The wavelet spectra at this instant of time has a well defined V-shaped high energy region around $x=9.52m$, with significantly higher energy than other locations. Additionally, the V-shape is well-defined since there is energy in the lowest scales of the wavelet transform at this location. It is known that as the scale of the wavelet transforms becomes smaller, the wavelet becomes more compressed. Therefore when a rogue wave occurs, which typically has a high and narrow peak, the small wavelet transform scales exhibit energy. This behavior can be observed in the Figure~\ref{fig8} and in the Figure~\ref{fig9} where the 1$^{st}$, 3$^{rd}$ and 8$^{th}$ order wavelet scales are presented.

\begin{figure}[htb!]
\begin{center}
   \includegraphics[width=3.4in]{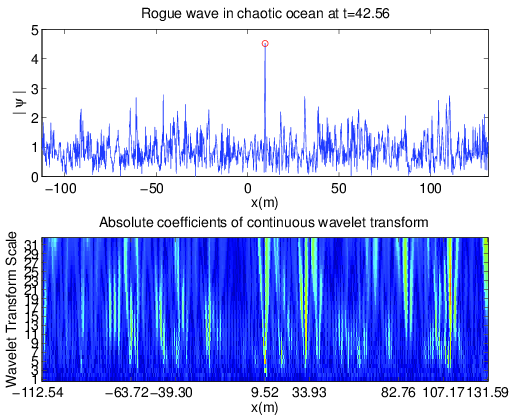}
  \end{center}
\caption{\small a) Rogue wave in chaotic wave field b) its wavelet transform, t=42.56s.}
  \label{fig8}
\end{figure}

\newpage

\begin{figure}[h]
\begin{center}
   \includegraphics[width=3.4in]{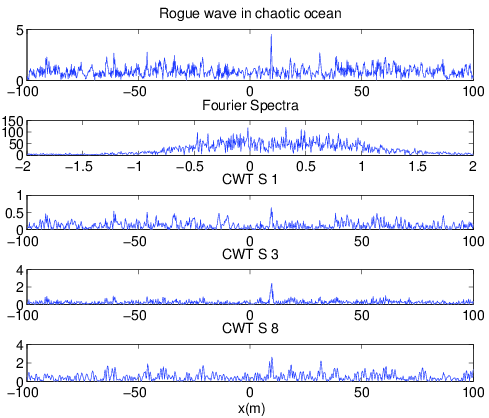}
  \end{center}
\caption{\small a) Rogue wave in the chaotic wave field, b) its Fourier transform c)-d)-e) its wavelet transforms of $1^{st}$, $3^{rd}$ and $8^{th}$ scale, respectively, t=42.56s.}
  \label{fig9}
\end{figure}

\begin{figure}[h]
\begin{center}
   \includegraphics[width=3.4in]{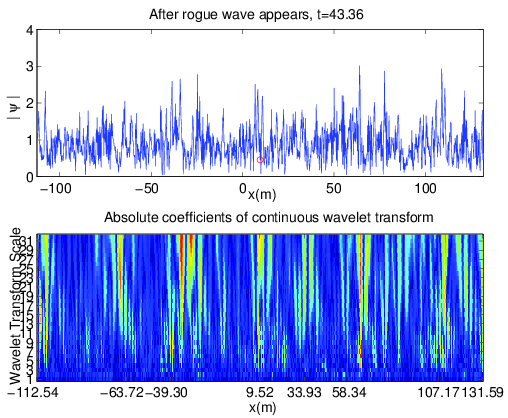}
  \end{center}
\caption{\small a) Chaotic wave field after rogue wave disappears b) its wavelet transform, t=43.36s.}
  \label{fig10}
\end{figure}

The Figure~\ref{fig10} shows the chaotic field after the rogue wave has disappeared. The wavelet spectra at this instant of time has still a well defined V-shaped high energy region around $x\approx 10m$ and another V-shaped high energy region around $x\approx 60m$. The difference between these locations leads to a value of $x\approx 50m$ which is the twice of the characteristic distance calculated previously for this chaotic wave field.

In reality, the accurate predictions of the rogue waves would include many factors \cite{Akhmediev2011}. These include but are not limited to the sensitivity and sampling rate of the measurement devices, the area chosen for the wavelet analysis, noise etc. \cite{Akhmediev2011}. In order to examine various nonlinearity and dispersion effects on rogue wave generation and early warning, the analysis offered in this study can be applied to other dynamic equations including but are not limited to generalized and high order NLSE, derivative NLSE \cite{Xu2011, Xu2012}, Kundu-Eckhaus equation \cite{Qiu2015} etc. Although the prediction of the rogue wave distance (time) be insufficient for with the current tools and technology, our suggestion of using the wavelet transforms for the early detection of the rogue waves seems to be promising and more advantageous than using the triangular Fourier spectra solely. As discussed above using wavelet transforms over (or together with) the Fourier transforms has several advantages such as predicting the location and characteristic distances between rogue waves. Our simple examples show that wavelet spectra have a specific V-shaped high energy area before the appearance of the high and rogue waves in spatial domain. In our future research we plan to extend our analysis to early warning and post disaster assessment of multi-rogue wave emergence such as those discussed in \cite{dubard2013multi, Kedziora2011, He} analytically.

\section{Conclusion}
In this paper we have discussed the possible advantages of using the wavelet transform over the Fourier transform for the early detection of rogue waves. It is well known that wavelet transforms preserves the spatial (temporal) information while calculating the spectra. Therefore compared to the Fourier spectra, the wavelet spectra can be used to predict not only the occurrence but also possible spatial (or temporal) locations of the rogue waves. Therefore the wavelet transform is also capable of predicting the distances between rogue and high waves which can turn out to be successive rogue waves (such as the famous ``three-sisters''). This result can enhance the safety of the marine travel and offshore structures and operations by improving the rogue wave early-warning systems.




\end{document}